\documentstyle[eqsecnum,aps]{revtex}

\begin{document}
\draft \preprint{HEP/123-qed}
\title{Particle Path Correlations in a Phonon Bath}
\author{Marco Zoli}
\address{Istituto Nazionale Fisica della Materia - Universit\'a di Camerino, \\
62032 Camerino, Italy. e-mail: marco.zoli@unicam.it }

\date{\today}
\maketitle
\begin{abstract}
The path integral formalism is applied to derive the full
partition function of a generalized Su-Schrieffer-Heeger
Hamiltonian describing a particle motion in a bath of oscillators.
The electronic correlations are computed versus temperature for
some choices of oscillators energies. We study the perturbing
effect of a time averaged particle path on the phonon subsystem
deriving the relevant temperature dependent cumulant corrections
to the harmonic partition function and free energy. The method has
been applied to compute the total heat capacity up to room
temeperature: a low temperature upturn in the heat capacity over
temperature ratio points to a glassy like behavior ascribable to a
time dependent electronic hopping with variable range in the
linear chain.
\end{abstract}

\pacs{PACS:71.38.+i, 31.15.Kb }

\section*{1. Introduction}

Determining the polaron properties is a relevant problem in many
body theory \cite{rash,devreese}. The electron motion through the
crystal is generally accompanied by a lattice deformation whose
size and entity depend both on the strength of the electron-phonon
coupling and on the value of the adiabaticity parameter peculiar
of the system
\cite{eminholst,firsov,toyo,der,alex,fehs,kopi,korni,jeck,mel,chr,rome,io,voul,trug,misc}.
In general, the lattice deformation does not follow
instantaneously the electron dragging it and the  retardation
effect becomes a key ingredient of this many body problem. The
Holstein \cite{holst} and the Su-Schrieffer-Heeger (SSH)
\cite{ssh} Hamiltonian models are fundamental tools in polaron
physics. While the former has been mainly used to describe
electron coupling to local optical phonons, both acoustic
\cite{ono} and optical branches are present in the dimerized
lattice of the latter. However, acoustic phonons in the SSH model
are not essentially affected by the e-ph coupling whereas optical
phonons are strongly softened due to single electron polarization
\cite{naka}.

In this paper we address the general problem of the interplay
between an electron and the optical phonon subsystem (a set of
independent oscillators providing the bath for the particle) in a
generalized SSH model, studying i) the temperature dependent
electronic correlations induced by the dissipative phonon bath and
ii) the thermodynamical behavior of the latter in the presence of
a perturbating particle motion. The path integral method
\cite{feynman}, being valid for any value of the {\it e-ph}
coupling, seems particularly suitable to our task. Moreover, it
naturally introduces the time (as an inverse temperature) into the
formalism thus allowing us to derive the thermal particle
correlation functions and the cumulant expansion for the phonon
free energy. In Section 2, we present the model and discuss the
particle correlations in the oscillators bath. The phonon free
energy is calculated in Section 3 while in Section 4 we apply the
model to compute the total heat capacity of the system both in
weak and strong {\it e-ph} coupling regimes. Some conclusions are
drawn in Section 5.

\section*{2. The Model}

The 1D SSH interacting Hamiltonian reads:

\begin{eqnarray}
H=\,& & \sum_{r}J_{r,r+1} \bigl(f^{\dag}_r f_{r+1} +
f^{\dag}_{r+1} f_{r} \bigr) \,
 \nonumber \\
J_{r,r+1}=\,& & - {1 \over 2}\bigl[ J + \alpha (u_r -
u_{r+1})\bigr] \label{1}
\end{eqnarray}

where $J$ is the hopping integral for an undistorted chain,
$\alpha$ is the electron-phonon coupling, $u_r$ is the
dimerization coordinate relative to the displacement of the atomic
group on the $r-$ lattice site along the molecular axis,
$f^{\dag}_r$ and $f_{r}$ create and destroy electrons on the $r-$
group. The free Hamiltonian is given by a set of classical
independent oscillators. By introducing $x(\tau)$ and $y(\tau')$
as the electron coordinates at the $r$ and $r+1$ lattice sites,
respectively and mapping $ u_r \to u(\tau)$ and $u_{r+1} \to
u(\tau')$ we transform the real space Hamiltonian of eq.(1) into
the time dependent Hamiltonian:

\begin{eqnarray}
H(\tau,\tau')=\,& &  J_{\tau, \tau'} \Bigl(f^{\dag}(x(\tau))
f(y(\tau')) + f^{\dag}(y(\tau')) f(x(\tau)) \Bigr)\, \nonumber \\
J_{\tau, \tau'}=\,& & - {1 \over 2}\bigl[J + \alpha(u(\tau) -
u(\tau'))\bigr] \label{2}
\end{eqnarray}

The SSH Hamiltonian has a twofold degenerate ground state which
undergoes a Peierls instability. In real space the soliton
connects the two degenerate phases with different senses of
dimerizations and a localized electronic state is associated with
each soliton. Both electron hopping to band states (thermal
excitation) and electron hopping between solitons are allowed.

Mapping the Hamiltonian onto the time scale we set up the finite
$T$ formalism in which thermally activated electron hops become
time dependent and the retarded nature of the interactions is
accounted for. $H(\tau,\tau')$ is  more general than the real
space SSH Hamiltonian since hopping processes are not constrained
to first neighbors sites along the chain. A variable range hopping
introduces some local disorder in the system \cite{lu}, therefore
the present approach may apply to polymers in which hopping
conduction mechanisms prevail. We point out that in real systems,
the Peierls gap can be smeared by temperature or doping induced
disordering effects \cite{rice}. Eq.(2) displays the semiclassical
nature of the model as quantum mechanical degrees of freedom
interact with the classical variables $u(\tau)$. Setting
$\tau'=\,0$, $u(0)\equiv y(0) \equiv 0$, averaging the electron
operators over the ground state and pinning the chemical potential
to the zero energy level, we write the average energy per lattice
site:

\begin{eqnarray}
& &{{<H(\tau)>} \over N}=\, V\bigl(x(\tau)\bigr) +
u(\tau)j(\tau)\,
 \nonumber \\
& &V\bigl(x(\tau)\bigr)=\,-J {a \over \pi} \int_{0}^{\pi/a}{dk}
\cos[k x(\tau)]  \cosh(\epsilon_k \tau /\hbar) n_F(\epsilon_k) \,
\nonumber \\ & &j(\tau)= -\alpha {a \over \pi}
\int_{0}^{\pi/a}{dk} \cos[k x(\tau)]  \cosh(\epsilon_k \tau
/\hbar) n_F(\epsilon_k) \, \nonumber \\ \label{3}
\end{eqnarray}

where $N=\,L/a$, with $L$ the chain length and $a$ the lattice
constant. $n_F$ is the Fermi function and $\epsilon_k=\,
-J\cos(k)$ is the electron dispersion relation.
$V\bigl(x(\tau)\bigr)$ is an effective term accounting for the
$\tau$ dependent electronic hopping while $j(\tau)$ is the
external source current for the oscillator path $u(\tau)$.
Averaging the electrons over the ground state we neglect the
fermion-fermion correlations \cite{hirsch} which lead to effective
polaron-polaron interactions in non perturbative analysis of the
model. This approximation however is not expected to affect
substantially our thermodynamical calculations. Being the energy
in eq.(2) linear in the displacements we can write the general
path integral \cite{kleinert} at any temperature as:

\begin{eqnarray}
& &<x(\beta)|x(0)>=\,\prod_i \int Du_i(\tau) \int Dx(\tau) \cdot
\, \nonumber \\ & &exp\Biggl[-{1 \over \hbar} \int_0^{\hbar \beta}
d\tau \sum_i {{M_i} \over 2} \Bigl(\dot{u_i}^2(\tau) + \omega_i^2
u_i^2(\tau) \Bigr) \Biggr] \cdot \, \nonumber \\ & &exp\Biggl[-{1
\over \hbar} \int_0^{\hbar \beta} d\tau \biggl({m \over 2}
\dot{x}^2(\tau) + V\bigl(x(\tau)\bigr) - \sum_i
u_i(\tau)j(\tau)\biggr) \Biggr] \, \nonumber
\\ \label{4}
\end{eqnarray}

where we have taken a large number of oscillators ($u_i(\tau),
i=1..{\bar N}$) as the {\it bath} for the quantum mechanical
particle whose coordinate is $x(\tau)$. $\beta$ is the inverse
temperature, $m$ is the electron mass and $\omega_i$ are the
oscillators frequencies. The oscillator masses are considered as
independent of $i$, $M_i\equiv M$ and hereafter we set
$M=\,10^4m$. After integrating out the oscillators coordinates
over the paths $Du_i(\tau)$, imposing a closure condition
$\Bigl(x(\beta)=\,x(0)\Bigr)$ on the particle paths and replacing
$\tau \to \tau/\hbar$, we obtain the full partition function in
the functional form

\begin{eqnarray}
& &Z(j(\tau))=\,Z_{ph} \oint Dx(\tau)exp\Bigl[-{m \over 2}
\dot{x}^2(\tau) - V\bigl(x(\tau)\bigr) - A(j(\tau)) \Bigr] \,
\nonumber \\ & &Z_{ph}=\, \prod_{i=1}^{\bar N} {1 \over
{2\sinh(\hbar\omega_i\beta/2)}} \, \nonumber \\ & &A(j(\tau))=\,
-{{\hbar^2} \over {4M}}\sum_{i=1}^{\bar N} {1 \over {\hbar
\omega_i \sinh(\hbar\omega_i\beta/2)}}\cdot \int_0^{\beta} d\tau_1
j(\tau_1) \int_0^{\beta}d{\tau_2} \cosh\Bigl(\omega_i \bigl(
|\tau_1 - {\tau_2}| - \beta/2 \bigr) \Bigr) j({\tau_2}), \,
\nonumber
\\
\label{5}
\end{eqnarray}

The thermodynamics of the interacting system can be derived from
eqs.(5) as a function of $J$, $\alpha$ and oscillators bath. An
application  is shown in Section 4 where the total heat capacity
is computed versus temperature. In terms of the generating
functional $Z(j(\tau))$ the two particle correlation function is
defined as

\begin{equation}
G^{(2)}(\tau_1, \tau_2)=\, \hbar^2 \Biggl[ Z^{-1}(j) {{\delta^2}
\over {\delta j(\tau_1) \delta j(\tau_2)}} Z(j) \Biggr]_{j=0}
\label{6}
\end{equation}

Then, using eqs.(5) and the fact that the source action
$A(j(\tau))$ is quadratic in the current, we can study the effect
of the phonon bath on the electronic time correlations at any
temperature. Setting $\tau_2=0$, we plot in Figures 1 the square
root of $G^{(2)}(\tau) \equiv <x(\tau) x(0)>$ ($\tau \in
[0,\beta]$) for three choices of oscillator baths: i) a {\it low}
phonon spectrum made of ten oscillators $\hbar \omega_{1}=\,2meV
..... \hbar \omega_{10}=\,20meV$ (spaced by $2meV$), ii) an {\it
intermediate} phonon spectrum with $\hbar \omega_{1}=\,22meV .....
\hbar \omega_{10}=\,40meV$ and iii) an {\it high} phonon spectrum
with $\hbar \omega_{1}=\,42meV ..... \hbar \omega_{10}=\,60meV$.
$\bar N$ and oscillator energies are purely arbitrary. The
periodic condition $G^{(2)}(\tau)=\,G^{(2)}(\beta - \tau)$ holds
as a general property of the particle-particle correlation
function which, up to a prefactor, coincides with the sum of the
oscillators Green functions and therefore it does not depend on
the width of the electron band. At low temperatures (Fig.1(a)),
small oscillator energies favor larger electronic correlations
which however get the maximum value of $0.26 \AA$ at $\tau
\rightarrow 0$. Increasing the temperature at $T=\,100K$
(Fig.1(b)) leads to a substantial enhancement of
$\sqrt{G^{(2)}(\tau)}$ which becomes of order $1 \AA$ in the case
of a low energy phonons bath. At room temperature (Fig.1(c)) and
for the same bath the electrons are correlated over a distance of
about $2 \AA$ in the whole $\tau$-range. Larger phonon spectra
tend to lock the electrons thus reducing the particle mobility
over the time scale and decreasing the space electronic
correlations. Since the upper limit in the $\tau$ range of Figs.
1(a)-1(c) is necessarily $T$ dependent we plot in Fig.2, for an
{\it intermediate} phonon bath, $\sqrt{G^{(2)}(\tau)}$ versus a
{\it common $\tau$-axis} to emphasize the temperature effect on
the particle correlations. Only the room temperature correlation
function retains periodicity over the selected $\tau$-range.

While we focus in the following on the equilibrium thermodynamics
of the particle-phonons interacting system, it should be noted
that generalization of eq.(4) through the closed-time path
formalism \cite{schwinger} would permit to derive dissipative
properties due to the phonon bath friction from the two points
correlation function. A density matrix study for the similar
problem of a particle in a photon bath (in three dimensions) has
been carried out by Haba and Kleinert \cite{haba}.

\section*{3. Phonon Free Energy}

Next we turn to analyse the effect of the particle motion on the
phonon subsystem. The set of $\bar N$ independent oscillators is
perturbed by the particle path $x(\tau)$ which couples to each
oscillator through the strength $\alpha$ (assumed independent of
$i$) of the SSH Hamiltonian. Then, the partition function of the
disturbed oscillators system can be expanded in perturbation
series as

\begin{eqnarray}
& &Z_{ph}[x(\tau)]\simeq \, \bigl(1 - <C> + <C^2> - <C^3> + \cdot
\cdot \bigr)Z_{ph}\, \nonumber \\ & &<C^k>=\,Z_{ph}^{-1} \prod_i
\oint Du_i(\tau) {{\alpha^k} \over {k!}} \int_0^{\beta} d\tau_1
u_i(\tau_1)x(\tau_1) \cdot \cdot \int_0^{\beta} d\tau_k
u_i(\tau_k) x(\tau_k)  \cdot \, exp\Biggl[- \int_0^{\beta} d\tau
\sum_i {M_i \over 2} \bigl( \dot{u_i}^2(\tau) + \omega_i^2
u_i^2(\tau) \bigr) \Biggr]\, \nonumber \\ \label{7}
\end{eqnarray}

In general, the total phonon partition function in presence of an
ensamble of particle paths $\bigl[x(\tau) \bigr]$ would be given
by: $Z^T_{ph}=\, \oint Dx(\tau) Z_{ph}[x(\tau)]$. Here we take a
single particle path, approximating it by the averaged
(dimensionless) value $<x(\tau)>\equiv\,{1 \over
\beta}\int_0^{\beta} d\tau x(\tau) =\,x_0$. Then, the odd k-terms
in the series expansion vanish. Since the oscillators are
decoupled (and anharmonic effects mediated by the particle path
are here neglected) we can study the behavior of the cumulant
terms $<C^k>$ by selecting a single oscillator having energy
$\omega$ and displacement $u(\tau)$. Hence, after expanding the
oscillator path in $N_F$ Fourier components

\begin{equation}
u(\tau)=\,u_o + \sum_{n=1}^{N_F} 2\Bigl(\Re u_n \cos( \omega_n
\tau) - \Im u_n \sin( \omega_n \tau) \Bigr)\, \label{8}
\end{equation}

with $\omega_n=\, 2n\pi/\beta$ and taking the measure of
integration

\begin{equation}
\oint Du(\tau)\equiv  \biggl({1 \over 2}\biggr)^{2N_F} {{\Bigl(
2\pi \cdot \cdot 2N_F \pi \Bigr)^2} \over {\sqrt{2}
\lambda_M^{(2N_F+1)} }} \int_{-\infty}^{\infty}{du_o}
\prod_{n=1}^{N_F} \int_{-\infty}^{\infty} d\Re u_n
\int_{-\infty}^{\infty} d\Im u_n \label{9}
\end{equation}

with $\lambda_M=\,\sqrt{\pi \hbar^2 \beta/M}$, we obtain for the
$k-th$ cumulant the following expression

\begin{equation}
<C^k>_{N_F}=\,Z^{-1}_{ph}{{\bigl( \alpha x_0 \bigr)^k} \over {k!}}
{{(\beta \lambda_M)^k (k - 1)!!} \over {\pi^{k/2} (\omega
\beta)^{k+1}}} \cdot {{(2\pi)^2} \over {(2\pi)^2 + (\omega
\beta)^2}} \cdot \cdot {{(2N_F \pi)^2} \over {(2N_F \pi)^2 +
(\omega \beta)^2}} \label{10}
\end{equation}

Since the cumulants should not depend on the number of Fourier
components in the particle path expansion, eq.(10) provides a
criterion to set the minimum $N_F$ (at any temperature and for any
oscillator) through the condition $2N_F \pi \gg \omega \beta$. The
thermodynamic properties of the perturbed oscillator can be
computed observing that the cumulant corrections to the free
energy are given by:

\begin{equation}
F_{cum}=\,-{1 \over \beta} \ln \Bigl[1 + <C^2> + <C^4> + <C^6> +
.. \Bigr] \label{11}
\end{equation}

In Fig.3(a), the three lowest order cumulants are plotted for a
weak {\it e-ph} coupling and an high frequency oscillator. At very
low temperatures, the cumulants attain the largest values which
are substantially independent of the $k$-order. By increasing $T$
all cumulants decrease, $<C^2>$ yields the dominant contribution
and the perturbation expansion converges rapidly. As shown in
Fig.3(b), the free energy corrections are not relevant due to the
weak $\alpha$ coupling. Note that $F_{cum}^{(k)}$ includes
$F_{ph}$ {\it plus} the perturbation given by eq.(11) to the
$k$-order. The $k=\,4$ correction is not appreciable with respect
to $F_{cum}^{(2)}$.

In Figures 4 we take a sizeable {\it e-ph} coupling (whose order
of magnitude applies to a polymer as polyacetylene), a factor ten
larger than in Figs.3: at low $T$, high order cumulants have to be
included in the series expansion but the temperature range (in
which their contribution is relevant) shrinks by increasing $k$.
An high number of Fourier components is required in order to get
numerical convergence (i.e., $N_F \sim 18000$ at $T=\,1K$)
signalling that the particle dynamics strongly interferes with the
oscillator at low $T$. For any $\alpha$ and any $T$, one
determines the $k$-order which makes the cumulant series
convergent.  For instance, being at $T \sim 150K$: $<C^8> \prec
<C^6> \sim <C^4>$, the series can be truncated for $k=\,6$ at $T >
150K$. Instead, at $T \sim 90K$, $<C^{10}> \prec <C^8> \succ
<C^6>$ so that the $k=\,8$ term suffices in the series expansion
at larger temperatures. Note however that high order cumulant
terms should be handled with care in computation of physical
quantities since they may provide some significant contributions
to the free energy, mainly at growing temperatures. As it is shown
in Fig.4(b), the largest correction to the free energy is due to
the $k=\,2$ term which strongly reduces the harmonic value in the
whole temperature range whereas the $k=\,4$ and $k=\,6$
contributions are also relevant although of decreasing importance.
The $k=\,8$ correction, still appreciable in the range $50K
\preceq T \preceq 200K$, tends to vanish above room temperature.
By enhancing the temperature, cumulants and free energy become
numerically stable taking a smaller $N_F$: at $T=\,300K$, $N_F
\sim 60$.

\section*{4. Heat Capacity}

By mapping the electronic hopping motion onto the time scale, we
have introduced a continuum version of the interacting SSH
Hamiltonian . Unlike previous \cite{lu} approaches however, our
path integral method is not constrained to the weak {\it e-ph}
coupling regime and it can be applied to any range of physical
parameters. To compute eqs.(5) one has to select the class of
particle paths which mainly contribute to the partition function
and fix the physical quantities characterizing the system: the
bare hopping integral $J$, the oscillator frequencies $\omega_i$
and the effective coupling $\chi=\, \alpha^2 \hbar^2 /M$ (in units
$meV^3$).

We take here a narrow band system ($J=\,100meV$) to be consistent
with previous investigations \cite{io1} and with the caveat that
electron-electron correlations may become relevant in narrow
bands. The total heat capacity has been first (Fig.5) computed up
to room temperature assuming the {\it low} phonon spectrum of
Fig.1. The lowest energy oscillator yields the largest
contribution to the phonon partition function mainly in the low
temperature regime while the $\omega_{10}$ oscillator essentially
sets the phonon energy scale which determines the size of the {\it
e-ph} coupling. A larger number $\bar N$ of oscillators in the
aforegiven range would not significantly modify the calculation.

In the discrete SSH model, the value $\bar \alpha \equiv \,
4\alpha^2/(\pi \kappa J) \sim 1$, marks the crossover between weak
and strong {\it e-ph} coupling, with $\kappa$ being the effective
spring constant. In the continuum and semiclassical model of
eq.(5) the effective coupling is the above defined $\chi$.
Although in principle, discrete and continuum models may feature
non coincident crossover parameters, we assume that the relation
between $\alpha$ and $J$ obtained by the discrete model crossover
condition still holds in our model. Hence, at the crossover we
get: $\chi_c \sim \, \pi J \hbar^2 \omega^2_{10}/64$. This means
that, in Fig.5, the crossover is set at $\chi_c \sim \,2000meV^3$.
The {\it total heat capacity over temperature} ratio shows a
peculiar low temperature upturn (also in the weak $\chi$ regime)
which can be mainly ascribed to the sizable effective hopping
integral term $V\bigl(x(\tau)\bigr)$. The {\it e-ph} coupling
however determines the shape of the low $T$ anomaly. The small
phonon contribution to the heat capacity is also reported on to
point out that the Dulong Petit value is achieved at $T \succeq
200K$.

The effect of the oscillators on the heat capacity is pointed out
in Fig.6 where we take the {\it intermediate} phonon bath with
energies: $\hbar \omega_1=\,22meV,..., \hbar \omega_{10}=\,40meV$.
Accordingly the crossover is set at $\chi_c \sim \,8000meV^3$ and
three plots out of five lie in the strong {\it e-ph} coupling
regime. The heat capacity  grows fast versus temperature at strong
couplings due to the source action contribution whereas the
presence of the low $T$ upturn in the {\it total heat capacity
over T} ratio is confirmed. Note that, due to the enhanced
oscillators energies, the phonon heat capacity saturates at $T
\sim 400K$.

Integrating eq.(5), we select, at any temperature, the ensamble of
particle paths over which the hopping potential
$V\bigl(x(\tau)\bigr)$ is evaluated. This ensamble is therefore
$T$ dependent. However, given a single set of path parameters one
can monitor the $V\bigl(x(\tau)\bigr)$ behavior versus $T$. It
turns out that the hopping decreases  by lowering $T$ but its
value remains appreciable also at low temperatures ($\preceq
20K$). Since the $d\tau$ integration range is larger at lower
temperatures, the overall hopping potential contribution to the
total action is relevant also at low $T$. Precisely this property
causes the anomalous upturn in the heat capacity linear
coefficient. Summing over a large number of paths is essential to
recover the correct thermodynamical behavior of the heat capacity
in the zero temperature limit.

Further investigation also reveals that the upturn persists both
in the extremely narrow ($J \sim 10meV$) and in the wide band ($J
\sim 1eV$) regimes. Our method accounts for a variable range
hopping on the $\tau$ scale which corresponds physically to
introduce some degree of disorder along the linear chain. This
feature makes our model more general than the standard SSH
Hamiltonian  with only real space nearest neighbor hops. Although
I am not aware of any other direct computation of specific heat in
the SSH model, hopping type mechanisms have been suggested
\cite{kivel} to explain the striking conducting properties of
doped polyacetylene at low temperatures. Since the specific heat
directly probes the density of states and integrating over $T$ the
{\it specific heat over T} ratio one can have access to the
experimental entropy, the method here presented may provide a new
approach to analyse the transition to a disordered state which
indeed exists in polymers \cite{nieu}. In this regard it is worth
remarking that glassy systems \cite{zeller} infact exhibit a low
$T$ upturn in the specific heat over $T$ ratio due to tunneling
states for groups of atoms providing a non magnetic internal
degree of freedom in the potential structure \cite{io91,io01}.

\section*{5. Conclusions}

We have developed a general path integral method to study the
interplay between electron motion and oscillators bath in a
semiclassical Su-Schrieffer-Heeger model. Using the fact that the
time dependent average energy is linear in the oscillator
displacement, we have analytically deduced the full partition
function as a functional of the {\it e-ph} source term and
computed the two particle correlations induced by the phonon baths
at some selected temperatures: at increasing temperatures the
electrons are correlated over larger distances. Then, we have
focussed on the perturbing effect of the electron particle motion
on the thermodynamics of a single oscillator. After expanding the
time dependent oscillator path in Fourier series, we have derived
the $k-$ order cumulant correction (eq.(10)) to the oscillator
partition function in the case of a time averaged electron
particle path. At decreasing temperatures an increasing number
($N_F$) of Fourier components in the oscillator path is required
to compute the cumulant terms with accuracy. $N_F$ is also a
growing function of the oscillator energy. The proposed method
permits to evaluate, for a given {\it e-ph} coupling, the cumulant
corrections to the harmonic partition function and free energy of
each oscillator at any temperature. While higher $k-$ order terms
become more relevant at low $T$ and they have to be included in
the partition function series expansion, the main corrections to
the harmonic free energy are instead ascribable to the lower
$k$-order cumulants and their effect is also evident at increasing
temperatures. Further refinements on the presented results may be
obtained by integrating the oscillator partition function over an
ensamble of $\tau$ dependent particle paths. Finally, the path
integral method has been applied to compute the heat capacity of
the system as a function of the {\it e-ph} coupling and the
oscillators energies. We find a peculiar upturn in the low
temperature plots of the heat capacity over temperature ratio
indicating that a glassy like behavior can arise in the linear
chain as a consequence of a time dependent electronic hopping with
variable range.

\begin{figure} \vspace*{12truecm} \caption{Square
root of the two-points correlation functions in units \AA. A bath
of ten phonon oscillators is considered, the largest phonon energy
being $\hbar \omega_{10}=\,40meV$ (intermediate phonon spectrum)
and $\hbar \omega_{10}=\,60meV$ (high phonon spectrum).  (a)
T=\,1K, (b) T=\,100K, (c) T=\,300K.}
\end{figure}

\begin{figure}
\vspace*{8truecm} \caption{Square root of the two-points
correlation functions in units $\AA$  plotted, at different
temperatures, in the range $\tau < 0.04meV^{-1}$. An intermediate
phonon spectrum is considered, the largest oscillator energy being
$\hbar\omega_{10}=\,40meV$.}
\end{figure}

\begin{figure}
\vspace*{10truecm} \caption{(a) Neperian logarithms of the first
three cumulants in eq.(10). (b) Phonon free energy and anharmonic
corrections due to the first three cumulants. The e-ph coupling
$\alpha$ is in units $eV \AA^{-1}$. $\omega$ is the phonon
energy.}
\end{figure}

\begin{figure}
\vspace*{10truecm} \caption{ As in Fig.3 but with a large e-ph
coupling $\alpha$.}
\end{figure}

\begin{figure}
\vspace*{8truecm} \caption{Total heat capacity over temperature
for three values of the effective coupling $\chi$ (in units
$meV^3$). A bath of ten phonon oscillators is considered, the
largest phonon energy being $\hbar \omega_{10}=\,20meV$. The
phonon heat capacity is also plotted.}
\end{figure}

\begin{figure}
\vspace*{8truecm} \caption{Total heat capacity over temperature
for five values of the effective coupling $\chi$ (in units
$meV^3$). The largest oscillator energy of the phonon bath is
$\hbar \omega_{10}=\,40meV$.}
\end{figure}


\begin{references}
\bibitem{rash}
Rashba E I 1982 {\it Excitons} edited by Rashba E I, Sturge M D,
(North-Holland, Amsterdam)
\bibitem{devreese}
Devreese J T 1996 {\it Encyclopedia of Applied Physics}, {\bf 14}
(VCH Publishers, NY) pp 383
\bibitem{eminholst}
Emin D and Holstein T 1976  Phys.\ Rev.\ Lett. {\bf 36} 323
\bibitem{firsov}
Firsov Y A, Kabanov V V, Kudinov E K, Alexandrov A S 1999 Phys.\
Rev.\ B {\bf 59} 12132
\bibitem{toyo}
Toyozawa Y 1961 Prog. Theor. Phys. {\bf 26} 29
\bibitem{der}
De Raedt H and Lagendijk A 1983 Phys.\ Rev.\ B {\bf 27} 6097;
ibid., 1984 {\bf 30} 1671
\bibitem{alex}
Alexandrov A S, Kabanov V V, Ray D K 1994 Phys.\ Rev.\ B {\bf 49}
9915
\bibitem{fehs}
Fehske H, R\"oder H, Wellein G, Mistriotis A, 1995 Phys.\ Rev.\ B
{\bf 51} 16582
\bibitem{kopi}
Kopidakis G, Soukoulis C M, Economou E N 1995 Phys.\ Rev.\ B {\bf
51} 15038
\bibitem{korni}
Kornilovitch P E and Pike E R 1997 Phys.\ Rev.\ B {\bf 55} R8634
\bibitem{jeck}
Jeckelmann E and White S R 1998 Phys.\ Rev.\ B {\bf 57} 6376
\bibitem{mel}
de Mello E V and Ranninger J 1998 Phys.\ Rev.\ B {\bf 58} 9098
\bibitem{chr}
Zolotaryuk Y, Christiansen P L, Rasmussen J J 1998 Phys.\ Rev.\ B
{\bf 58} 14305
\bibitem{rome}
Romero A H, Brown D W, Lindenberg K 1999 Phys.\ Rev.\ B {\bf 59}
13728
\bibitem{io}
Zoli M 2000 Phys.\ Rev.\ B {\bf 61} 14523
\bibitem{voul}
Voulgarakis N K and Tsironis G P 2001 Phys.\ Rev.\ B {\bf 63}
14302
\bibitem{trug}
Ku L C, Trugman S A, Bon\v{c}a J 2002 Phys.\ Rev.\ B {\bf 65}
174306
\bibitem{misc}
Mischchenko A S, Nagaosa N, Prokof'ev N V, Sakamoto A, Svistunov B
V 2002 Phys.\ Rev.\ B {\bf 66} 020301(R)
\bibitem{holst}
Holstein T 1959 Ann.\ Phys.\ (N.Y.) {\bf 8} 325
\bibitem{ssh}
Su W P, Schrieffer J R, Heeger A J 1979 Phys.\ Rev.\ Lett. {\bf
42} 1698;  Heeger A J, Kivelson S, Schrieffer J R, Su W P 1988
Rev.\ Mod.\ Phys. {\bf 60} 781
\bibitem{ono}
Miyasaka N and Ono Y 2001 J.Phys.Soc.Jpn. {\bf 70} 2968
\bibitem{naka}
Nakahara M and Maki K 1982 Phys.\ Rev.\ B {\bf 25} 7789
\bibitem{feynman}
Feynman R P 1955 Phys. Rev. {\bf 97} 660
\bibitem{lu}
Lu Y 1988 {\it Solitons and Polarons in Conducting Polymers}
(World Scientific, Singapore)
\bibitem{rice}
Rice M J and Mele E J 1981 Chem.Scripta {\bf 17} 121
\bibitem{hirsch}
Hirsch J E 1983 Phys.\ Rev.\ Lett. {\bf 51} 296
\bibitem{kleinert}
Kleinert H 1995 {\it Path Integrals in Quantum Mechanics,
Statistics and Polymer Physycs} (World Scientific Publishing,
Singapore)
\bibitem{schwinger}
Schwinger J 1961 J.Math.Phys. {\bf 2} 407
\bibitem{haba}
Haba Z and Kleinert H 2001 Eur. Phys. J. B {\bf 21} 553
\bibitem{io1}
Zoli M 2002 Phys.\ Rev.\ B {\bf 66} 012303
\bibitem{kivel}
Kivelson S 1981 Phys.\ Rev.\ Lett. {\bf 46} 1344
\bibitem{nieu}
Nieuwenhuizen Th M arXiv:cond-mat/9701044
\bibitem{zeller}
Zeller R C and Pohl R O 1971 Phys.Rev.B {\bf 4} 2029
\bibitem{io91}
Zoli M 1991 Phys.\ Rev.\ B {\bf 44} 7163
\bibitem{io01}
Zoli M 2001 Phys.\ Rev.\ B {\bf 63} 174301
\end{references}
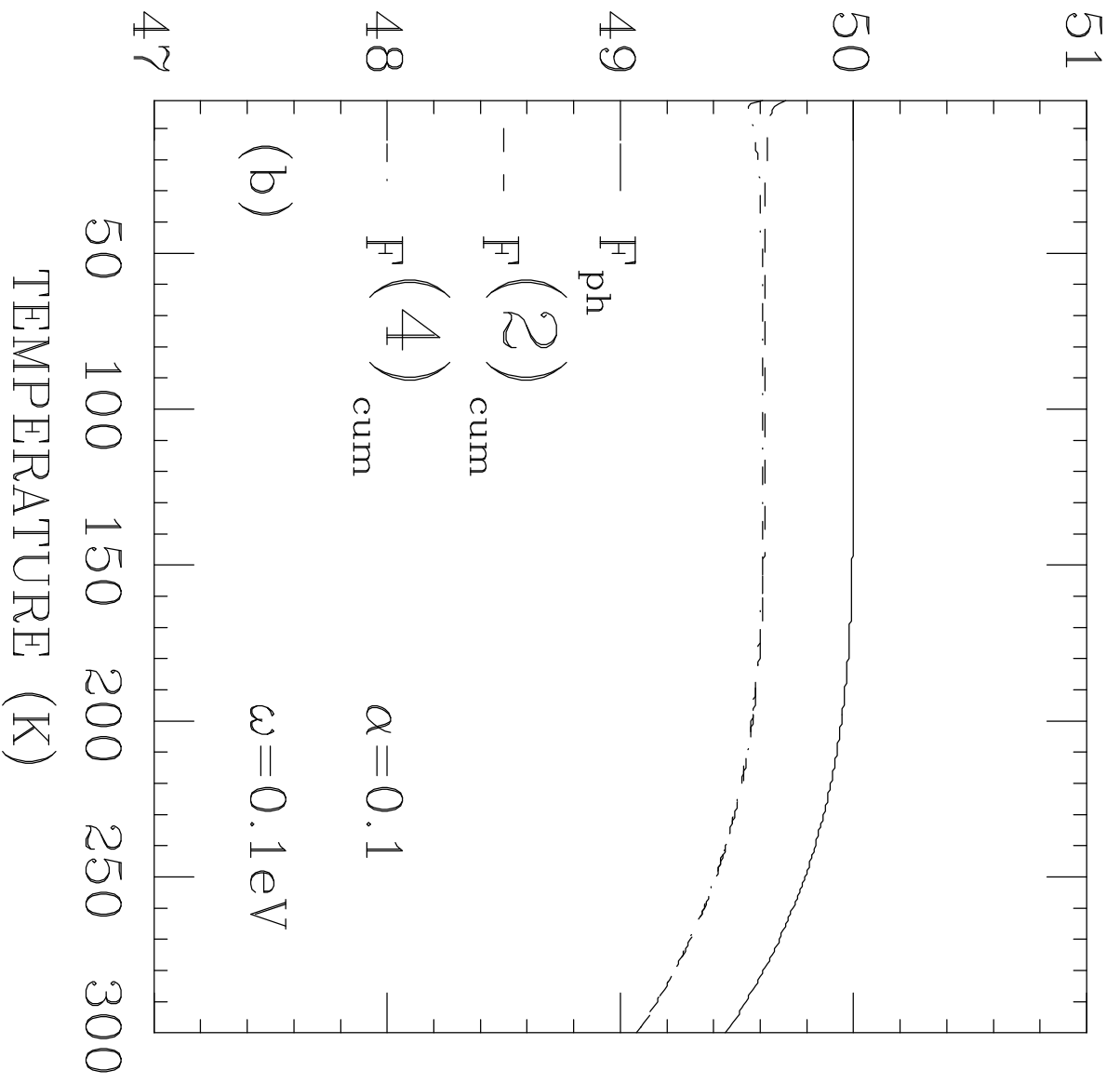
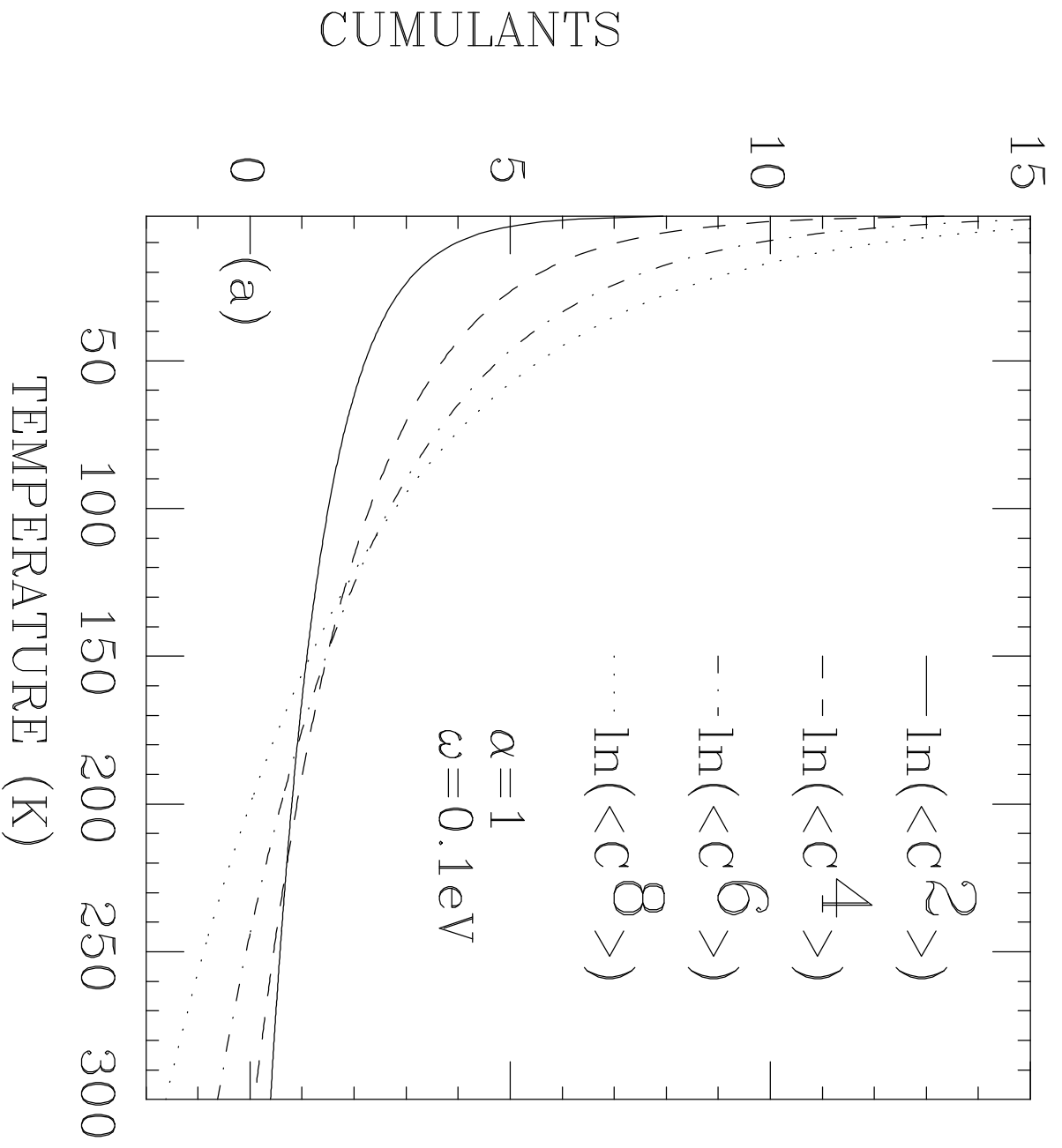
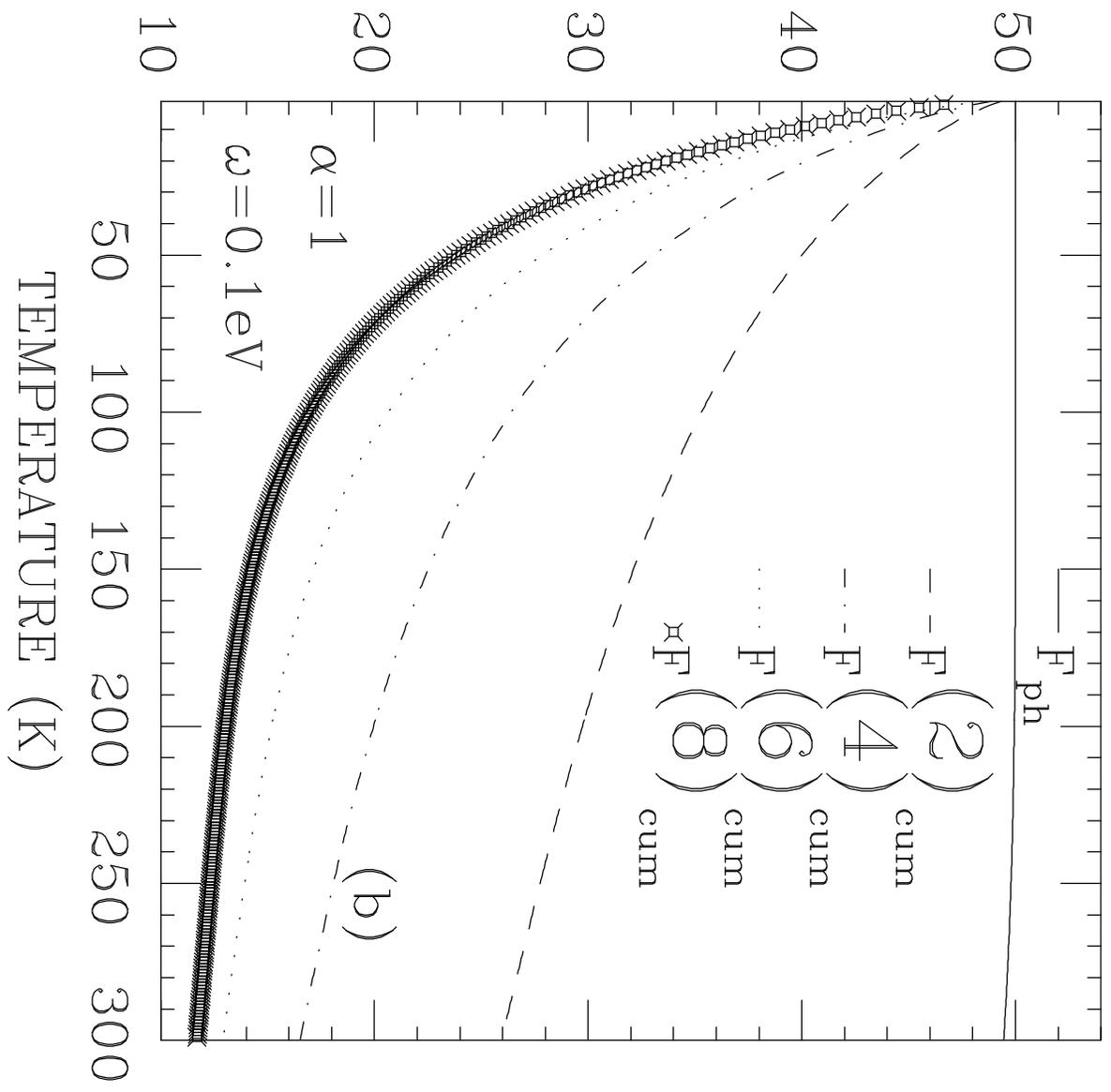
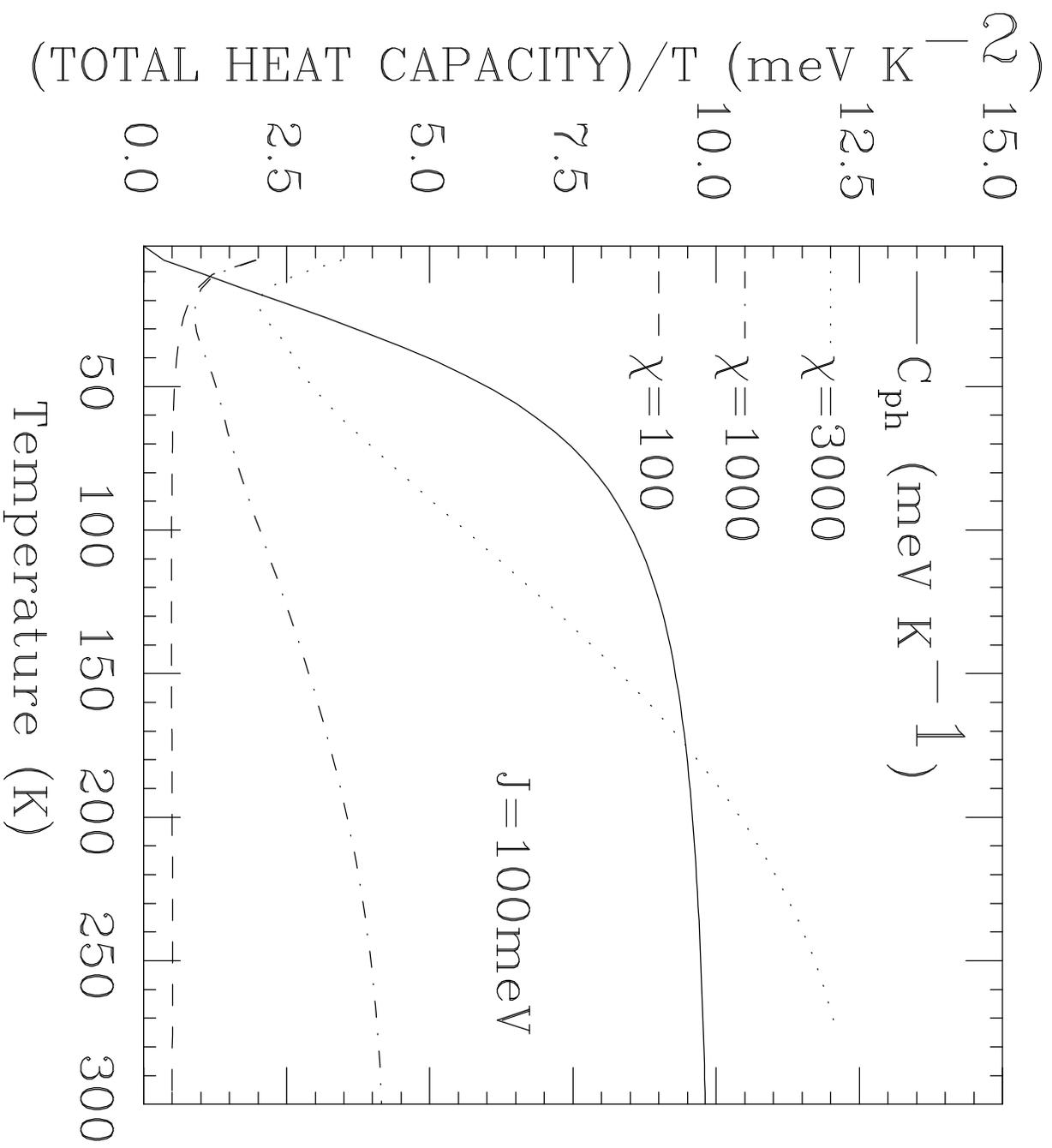
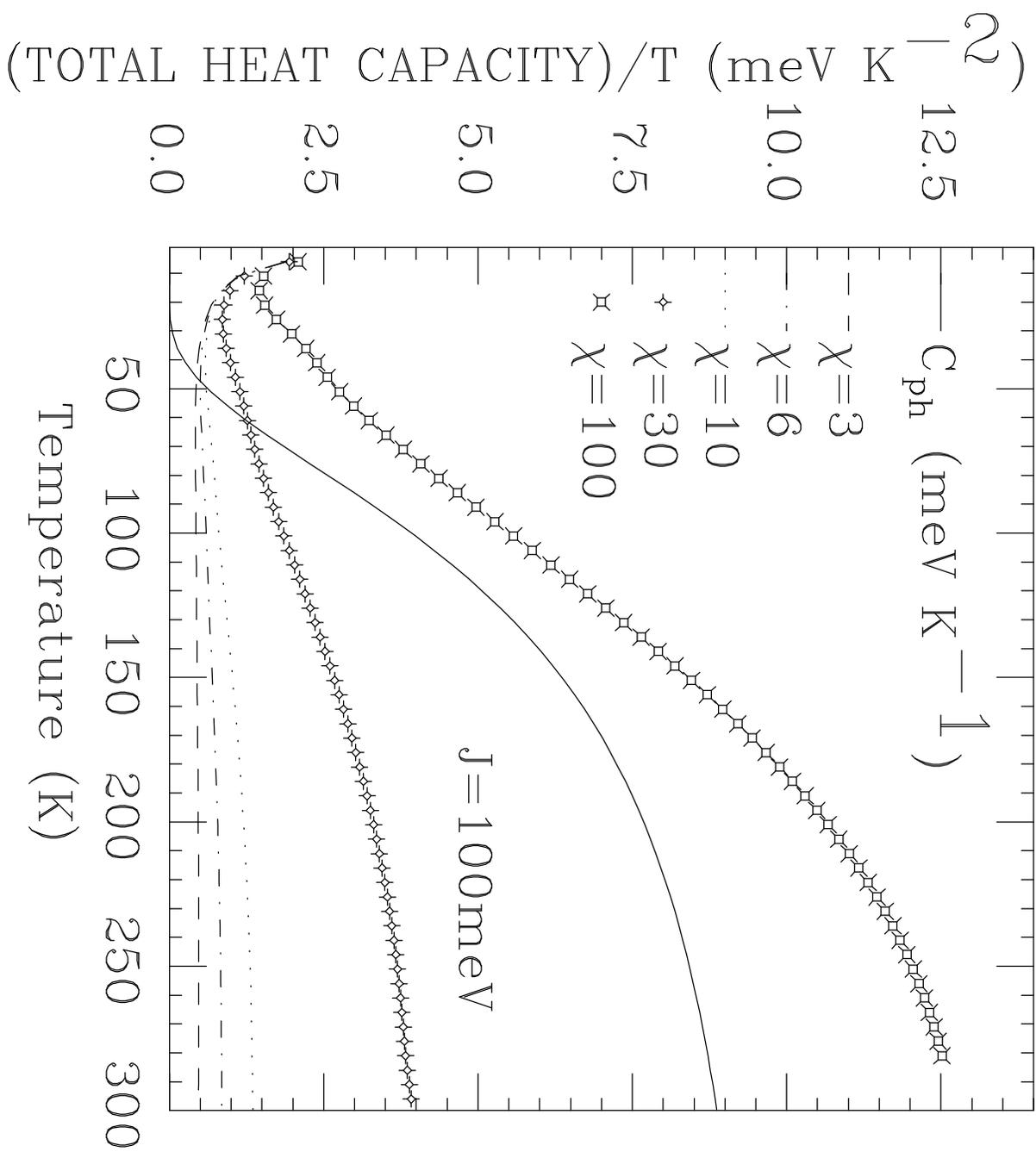
\end{document}